\documentstyle[aps,twocolumn]{revtex}
\begin{document}
\tighten
\draft
\title{Charged ``Few-Electron -- Single Spatially Separated Hole" 
Complexes in a Double Quantum Well near a Metal Plate}
\author{
V.I. Yudson\cite{Present}}
\address{Institute of Spectroscopy, Russian Academy of Sciences,
Troitzk, Moscow reg.~142092, Russia}
\date{Phys. Rev. Lett. {\bf 77}, No. 7}
\maketitle

\begin{abstract}
It has been shown that the presence of a metal plate near a double quantum
well with spatially separated electron and hole layers may lead to a 
drastic reconstruction of the system state with the formation of 
{\it stable charged} complexes of several electrons bound to a 
{\it spatially separated} hole. 
Complexes of both the Fermi and the Bose statistics may coexist in the 
ground state and their relative densities may be changed with 
the change of the electron and hole densities.
The stability of the charged complexes may be increased by an
external magnetic field perpendicular to the layers plane.
\end{abstract}

\narrowtext
Double layers (semiconductor double quantum wells (DQWs)) of spatially 
separated two-dimensional (2D) electrons ($e$) and holes ($h$) are a 
subject of considerable research interest. 
The Coulomb
$e$-$h$ interaction leads to interlayer correlations of the 
carriers and to the possibility of a transition to the superfluid 
excitonic phase \cite{LY,Sh}. 
There is a rapidly increasing amount of publications devoted to both 
theoretical \cite{CQ91,Li96,VM96}
and experimental \cite{Fuk90,But94}
investigations of e-h coupling in DQWs; see also references in 
\cite{Li96,VM96,But94}. Some 
experimental evidence has been reported \cite{But94}
for a stable excitonic ground state   
in a strong magnetic field which favors the stability of the excitonic 
phase \cite{LL}.

In contrast to the locally neutral ``excitonic insulator" 
phase of a bulk sample \cite{KK64}, the excitonic phase of 
the spatially separated $e$ and $h$ would not possess  
the {\it local} electrical neutrality and the motion of strongly coupled 
spatially separated charges would be accompanied by nonzero electric currents 
counterflowing on $e$- and $h$-layers \cite{LY}. However, these currents are 
of equal magnitudes, which means that the {\it total} current 
along the QWs plane equals zero. 

The question addressed in the present paper concerns an existence of 
stable {\it charged} bound complexes of few electrons and holes in the 
double-layer system with {\it non-equal} 2D densities $n_e$ and $n_h$ of 
spatially separated $e$ and $h$. 
The two simplest charged (quasi-)Fermi and (quasi-)Bose complexes 
(``electronic molecules") would be $e_2h$ and $e_3h$, respectively 
(for definiteness, we will assume the negative total charge).
The charged boson-like complexes would be of special interest due to 
the possibility of their Bose-Einstein condensation and superfluidity
accompanied with a nonzero {\it total} current along the DQW plane.

The fermion molecule $e_2h$ known as a negatively charged exciton 
($X^-$) does exist in bulk samples as well as in QWs and quantum dots.
A magnetic field favors the existence of the $X^-$ state 
(see, e.g. \cite{DS93,PYM96}),
and even a series of fermionic ``homologies" $X^-_K$ (i.e., $e_{K+1}h_K$)
with $K$ = 2,3,.. has been predicted
recenly \cite{PYM96}. 

However, no charged {\it boson} bound states of {\it few} $e$ and $h$ 
(not to be mixed with mesoscopic electron-hole droplets where the big 
numbers of $e$ and $h$ may differ slightly) have been found up to now. 
Their existence
is hindered because of the high increase of the Coulomb energy caused by 
{\it extra} electrons.
This concerns especially the systems of spatially separated $e$ and $h$ 
where the Coulomb repulsion of electrons dominates over their attraction 
to a distant hole. 

In this Letter we draw attention to a novel physical situation which takes
place in the DQW system of spatially separated $e$- and $h$-layers 
located near a parallel metal plate (MP). When the MP is
close to one of the layers (for definiteness, to the $e$-layer),
the $e$-$e$ Coulomb repulsion is suppressed considerably by
the ``image charge" polarization of the MP \cite{Im}.
The influence of a MP on the interaction and collective
properties of low-dimensional electron systems is not a new subject; it
is known, for instance, that the MP 
hinders the crystallization of the 2D electron gas
(``cold melting" of the 2D electron Wigner crystal \cite{LY75}).
However, to the best of our knowledge, the advantage of using this 
suppression of the $e$-$e$ repulsion in systems of spatially
separated $e$ and $h$ has not been studied yet. 

We demonstrate that the MP
close to the $e$-layer may lead to the formation 
of stable mobile charged complexes $e_Nh$ of several ($N$) electrons 
bound to a {\it spatially separated} hole. 
We consider $N$ electrons confined to the 2D layer and a single hole 
located at another parallel layer at a height $l$ over the $e$-layer.
The $e$-layer is at a height $d$ over the surface of a metal plate.
The potential energy of the $e$-$h$ system is given by 
\begin{eqnarray}\label{U}
U = &-& \sum_i[V({\bf r}_i-{\bf r}_h;l) - V({\bf r}_i-{\bf r}_h;l+2d)]
\nonumber \\
    &+& \frac{1}{2}\sum_{i \neq j}[V({\bf r}_i-{\bf r}_j;0) -
V({\bf r}_i-{\bf r}_j;2d)], 
\end{eqnarray}  
where ${\bf r}_i$ ($i = 1,..., N$) and ${\bf r}_h$ are 2D radius-vectors 
of the electrons and the hole, respectively; 
$V({\bf r};l) \equiv (e^2/\epsilon)[{\bf r}^2 + l^2]^{-1/2}$, 
$\epsilon$ is the dielectric constant of a surrounding medium.

Our present consideration is restricted to the case 
when $l$ is large as compared to the 
characteristic quantum lengths, which are the effective Bohr radii 
$a_{e,h}$ of $e$ and $h$
or the magnetic length $\lambda_H$
(if a strong magnetic field $H$ is applied
perpendicular to the DQW plane). In this limit, classical configurations
play the crucial role and determine the leading contribution to the system
energy. This is the feature of systems with spatially separated $e$ and 
$h$; in systems with no spatial separation of $e$ and $h$  
quantum mechanics is the only remedy against the collapse of 
the classical charges.
Surprizingly, it turns out that even the classical configurations of the
DQW-metal system are quite nontrivial.  
Later we will also discuss effects of quantum fluctuations.

For further comparison, first we describe 
classical 
configurations in the double-layer geometry without the MP (i.e., $d=\infty$).
We introduce the rectangular coordinate system ($\hat{x}$-$\hat{y}$ axes)
on the $e$-layer and set the coordinate origin $O$ exactly 
under the hole. The only stable classical state - ``$eh$" complex ($N = 1$) - 
corresponds to the electron located at the point $O$. The energy 
$U_1=- e^2/(\epsilon l)$ of this bound state defines the characteristic 
energy scale of the system. 
There exist also unstable equilibrium classical configurations $e_Nh$
with $2 \leq N \leq 4$ electrons located symmetrically at a ring 
centered at the $O$.
The ring radius $\rho$ is given by 
\begin{eqnarray}\label{R}
\rho = \tan{\theta},\,\,\,\,\,\, \theta = 
\arcsin{\left(\frac{1}{4}\sum^{N - 1}_{i = 1}{\frac{1}
{\sin{(\pi i/N)}}}\right)^{1/3}},
\end{eqnarray}  
which provides a saddle-point extremum of the potential energy 
(\ref{U}) at $2 \leq N \leq 4$.
Energies of these configurations are $U_2\approx-0.94 |U_1|$, 
$U_3\approx-0.51 |U_1|$, and $U_4\approx-3.3 \cdot 10^{-3} |U_1|$, 
respectively. These configurations
are unstable and decay into the stable $eh$ state and free 
electrons. Though the unstable
classical configurations
might manifest in nonstationary processes (e.g., optical absorption), 
they are not important in the equilibrium.
According to Eq.(\ref{R}), at $N > 4$ 
equilibrium classical configurations $e_Nh$ do not exist at all. 
To summarize, no {\it stable} classical configurations $e_Nh$ 
with $N > 1$ exist in the DWQ without a neigboring MP.

The presence of a MP at the distance $d$ down from the electron 
layer may change the situation qualitatively. This 
is particularly 
pronounced at $\eta \equiv d/l \ll 1$ and we begin with 
this range of parameters. 
The simplest $eh$ state ($N = 1$) corresponds to the electron located 
at the point $O$. The classical energy of this state is 
\begin{eqnarray}\label{U1}
U_1(\eta)/|U_1| = - 2\eta/(1 + 2\eta) \approx - 2\eta + O(\eta^2),
\end{eqnarray}  
where the second equality refers to the case $\eta \ll 1$. On the
contrary, with the increase of $\eta$, $U_1(\eta)$ tends to the energy 
$U_1 = - e^2/(\epsilon l)$ of the classical $eh$ configuration in the 
absence of the MP. 

At $\eta \ll 1$, the radius $\rho$ of the equilibrium classical 
configurations $e_2h$ 
obeys $d \ll \rho \ll l$ and
is determined as the extremum of the following approximate expression for the
potential energy Eq.(\ref{U}): $U/|U_1| \approx - 4\eta + 6\eta(\rho/l)^2 +
\eta^2 (l/\rho)^3/4$. We obtain 
$\rho
\approx 0.57 \eta^{1/5} l$ 
which justifies the approximations above. The classical energy of the
$e_2h$ configuration is given by
\begin{eqnarray}\label{U2}
U_2(\eta)/|U_1| = - 4\eta + 5\cdot2^{-3/5}\eta^{7/5} 
                \approx - 4\eta + 3.29 \eta^{7/5}. 
\end{eqnarray}  
Thus, due to the image charges which decrease the $e$-$e$ 
repulsion, the charged complex $e_2h$ has the lower energy (\ref{U2})
than the energy (\ref{U1}) of the neutral complex $eh$. The $e_2h$ complex
is the {\it stable classical ground state} of the ``two-electron - single 
hole" system. At the same time
$2U_1(\eta) < U_2(\eta)$, which guarantees the stability of the $eh$ 
complexes with respect to the reaction $2eh \rightarrow e_2h + h$.
This means that as far as the electron density $n_e$ does not exceed
the hole one $n_h$, all the electrons are bound into the $eh$ complexes.
(Here and below both $n_e$ and $n_h$ are assumed to be sufficiently low
so that we neglect all the screening effects; the temperature is 
also assumed to be sufficiently low). In the range 
$n_h < n_e < 2n_h$, boson $eh$ and fermion $e_2h$ complexes coexist and their 
densities equal $2n_h - n_e$ and $n_e - n_h$, respectively. 
The stability of complexes with respect to adhering to one another is
provided by the mutual hole repulsion which is less affected by 
relatively distant image charges.

For the simplest charged boson complex $e_3h$ we meet a new phenomenon
which has not occured in the absence of the MP - 
there are {\it two} possible classical equilibrium configurations 
$e_3h$-a and $e_3h$-b described as follows: 
a) 3 electrons form a regular triangle centered at $O$;
b) one of 3 electrons sits at the center $O$  (an ``inner shell")
and the others are located symmetrically with respect to the first one 
(an ``outer shell").
We obtain the 
radii 
$\rho \approx 0.72 \eta^{1/5}l$ and $\rho \approx 1.01 \eta^{1/5}l$ of  
$e_3h$-a and $e_3h$-b configurations, respectively, and the corresponding 
energies
\begin{eqnarray}\label{U3a}
U_{3a}(\eta)/|U_1| &   =   & - 6\eta + 5\cdot3^{2/5}\eta^{7/5}\nonumber \\
U_{3b}(\eta)/|U_1| &   =   & -6\eta + 10(17/16)^{2/5}\eta^{7/5}
\end{eqnarray}  
As $U_{3a}(\eta) < U_{3b}(\eta)$, the configuration 
$e_3h$-b is unstable with respect to the transition into the lower energy 
``isomer" configuration $e_3h$-a. Indeed, the stability analysis of $e_3h$-b
configuration reveals an unstable mode which 
tends to distort 
the electron configuration towards the triangle 
arrangement of the $e_3h$-a complex. 
As $U_{3a}(\eta) < U_{2}(\eta) < U_{1}(\eta)$ at small $\eta$, 
the charged boson complex $e_3h$-a realizes 
the classical ground state of the system of ``three electrons and one hole" 
(note in advance that the roles of $e_3h$-a and $e_3h$-b 
isomers will interchange when $\eta$ is not small).
At the same time, the inequalities $3U_{1}(\eta) < U_{3a}(\eta)$, 
$U_{1}(\eta) + U_{2}(\eta) < U_{3a}(\eta)$, and 
$2U_{2}(\eta) < U_{3a}(\eta) + U_1(\eta)$ forbid the reactions
$3eh \rightarrow e_3h\mbox{-a} + 2h$, $eh + e_2h \rightarrow e_3h\mbox{-a} + 
h$, and $2e_2h \rightarrow e_3h\mbox{-a} + eh$, respectively.
This means that the charged boson complexes $e_3h$-a may appear only at 
$n_e > 2n_h$. In the range 
$2n_h < n_e < 3n_h$, the boson $e_3h$-a and the fermion $e_2h$ complexes 
coexist; their densities equal $n_e - 2n_h$ and $3n_h - n_e$, respectively.

The problem one meets at higher $N$ is 
a variety of possible isomer configurations which correspond to different
arrangements of electrons over ``shells". 
The energies of type ``a" (the ``electron ring") and type ``b" 
(with an electron at the center of the electron ring) 
configurations at $\eta \ll 1$ are given by
\begin{eqnarray}\label{UNb}
&&U_{N\sigma}(\eta)/|U_1| = - 2N\eta \nonumber \\
&& + 5N\left[k_{\sigma} + \frac{1}{16}\sum^{N-1-k_{\sigma}}_{i=1}
{\frac{1}{\sin^3{[\pi i/(N-k_{\sigma})]}}}\right]^{2/5}\eta^{7/5},
\end{eqnarray} 
where $k_{\sigma} = 0$ and $k_{\sigma} = 1$ for the ``a" and ``b" 
configurations, respectively.
At $N > 5$ the configuration ``b" 
becomes a lower energy than the configuration ``a". It is not clear yet
whether there are even more favorable configurations. Complete consideration 
of a ``Periodic" Table of Complexes remains a subject for further
research. Here we only estimate 
the maximal allowed value of $N$ (i.e., the number $N_c$ of stable 
classical complexes in the Table) at a given $\eta$. 
A lower bound for this value is determined by the 
first violation of the condition $U_N(\eta) < U_{N-1}(\eta)$, 
which reduces to 
$\partial U_N(\eta)/\partial N = 0$ for large
$N$. If the ground state corresponds to 
the type ``b" configuration, we obtain the following estimate for $N_c$:
\begin{eqnarray}\label{Nc}
N_c = C/\eta^{1/3} \gg 1,
\end{eqnarray} 
where 
$C = 2\pi(2/11)^{5/6}$. If other configurations 
become important, 
the functional dependence Eq.(\ref{Nc}) might be
still valid although the numerical factor $C$ might change.
According to Eq.(\ref{Nc}), at small $\eta$ the world of stable 
charged complexes may be rather rich. However, their binding energies
at small $\eta$ are small, which is not favorable
for experimental realizations.

When $\eta$ increases (i.e., the metal plate is removed from the DQW),
the binding energies of possible stable configurations increase
but the number of these configurations decreases.
Our further consideration is 
restricted to the representative configurations $eh$, $e_2h$, $e_3h$-a, and 
$e_3h$-b.
Their energies in the intermediate range of $\eta$ are plotted as functions 
of $\eta$ on Fig.1. 
In the interval $0 < \eta < 1.32$ 
the fermionic complexes $e_2h$ are stable ($U_2(\eta) < U_1(\eta)$) and 
therefore they would be present at the system ground state at 
$n_h < n_e \leq 2n_h$. As to the charged boson complexes 
(at $2n_h < n_e \leq 3n_h$), they are present in the $e_3h$-a isomeric
configuration at $0 < \eta < 0.39$. At $\eta \approx 0.39$  
the isomer configuration $e_3h$-b becomes more favorable. 
The latter remains stable in the interval $0.39 < \eta < \eta_c = 16$ but
it does not exist even as an unstable equilibrium state at $\eta > \eta_c$.
The existence of the critical value of $\eta$ is a consequence of the 
fact that no equilibrium configuration of type ``b" 
may exist in the DQW
without the MP. As follows from Eq.(\ref{U}), at $d/l = \eta \rightarrow 
\eta_c - 0$ the radius of the $e_3h$-b configuration tends to infinity:
$\rho/l \approx 194/\sqrt{(16 - \eta)}$; correspondingly, the energy of
loosing the outer electron shell (i.e. the ``ionization" energy) decreases
drastically: $U_1(\eta) - U_{3b}(\eta) \propto (16 - \eta)^{5/2}|U_1|$.

A succession of the formation of charge complexes with the increase of the
electron density differs for different intervals of $\eta$. At 
$0 < \eta < 1.32$, this succession ($eh \rightarrow e_2h \rightarrow 
e_3h$) is similar to one described for small $\eta$ (with the replacement
of the a-isomer by the b-isomer at $\eta \approx 0.39$). However, at 
$1.32 < \eta$ the fermionic $e_2h$ complexes are not stable, and 
at $n_h < n_e \leq 3n_h$ {\it only boson} complexes $eh$ (with the density 
$3n_h/2 - n_e/2$) and $e_3h$-b (with the density $n_e/2 - n_h/2$) 
may coexist at the ground state of the low density $e$-$h$ system
at $1.32 < \eta < \eta_c = 16$.
At $\eta > 16$ only the neutral $eh$ complexes may exist in the ground 
state and the rest of electrons (at $n_e > n_h$) remain free.

To increase the typical energy scale $|U_1| = e^2/(\epsilon l)$ of the 
classical configurations it is desirable to decrease the interlayer 
distance $l$. However, this would increase quantum effects and eventually 
would make the classical description nonadequate.
The criterion of the classical approach validity is $\xi \ll \rho$, 
where $\xi \sim \sqrt{\hbar/(m\omega_0)}$ is an amplitude of ``zero" 
oscillations of the charges around their classical 
equilibrium positions and $\omega_0$ is a ``characteristic" frequency of
intracomplex oscillations. For the range of small $\eta=d/l$ and moderate
$N$, the oscillation frequencies differ only by numerical factors 
(for comparable values of $e$ and $h$ effective masses 
$m = m_e \sim m_h $) and scale as $\hbar\omega_0 \sim |U_1|da/l^2$
($a = \hbar^2/(me^2))$. This gives $\xi \sim l(a/d)^{1/4}$ and
determines the range of validity of the 
classical approach at small $\eta$:
\begin{equation}\label{d1}
a(l/a)^{4/9} \ll d \ll l.
\end{equation} 
In the intermediate range of $\eta$ (i.e. $d \sim l$) the situation is
more favorable. Typical intracomplex oscillation frequencies are now
estimated as $\omega_0 \sim [e^2/(ml^3)]^{1/2}$ and the requirement 
$\xi \ll \rho$ reduces to 
\begin{equation}\label{d2}
a \ll d \stackrel{<}{\sim} l,
\end{equation} 
which is weaker than Eq.(\ref{d1}). These estimations show that in the
well pronounced classical regime the typical classical energy scale $|U_1| = 
e^2/(\epsilon l)$ is considerably smaller than the exciton Rydberg. 
However, Eqs.(\ref{d1}) and (\ref{d2}) are only 
sufficient but not necessary conditions.
We may expect that the complexes survive even when the strong
left inequalities in Eqs.(\ref{d1}) and (\ref{d2}) are replaced by the 
usual ones.

Quantum effects may be suppressed by application of a strong 
magnetic field $H$ perpendicular to the layers plane. The magnetic field 
has no influence on the structure of classical configurations but it
induces a rotation of the electron ring (the Amp\`{e}re ``persistent" current)
and reduces the quantum ocsillation amplitudes of the charges. At 
$\lambda_H=\sqrt{\hbar c/(eH)}<\xi \sim l(a/d)^{1/4}$, the oscillation
amplitudes will be of the order of $\lambda_H$ and at sufficiently strong 
magnetic fields the condition $\lambda_H \ll \rho \sim l(d/l)^{1/5}$ 
provides the existence of stable classical charged configurations even at 
$a \sim d \stackrel{<}{\sim} l$, i.e. when the classical energies are 
comparable with the exciton Rydberg. 
It might also be favorable to use II - VI semiconductor DQWs where $a$ is
smaller than in currently used GaAs/AlGaAs and InAs/AlGaSb DQWs 
\cite{Fuk90,But94}.

To describe the most favorable range of parameters, the simplified 
quasiclassical consideration should be extended to the quantum one.
More elaborate study has to be done to describe the shell structure 
of this new kind of ``artificial atoms", to calculate the spectrum of their 
vibrational and rotational modes, and to fill the ``Table of Complexes".
Collective phenomena in the low-density system of complexes 
and, particularly, the tempting possibility of the Bose-Einstein condensation 
of the charged $e_3h$ ``bosons" also deserve a special research. 

The presence of the charged complexes may manifest 
in the Hall
conductivity and cyclotron resonance measurements, drag experiments, and
microwave absorption by intracomplex degrees of freedom.
Experimental search for the charged boson complexes $e_3h$ (or $eh_3$, if 
the MP is closer to the h-layer) of the most 
interest 
would be, perhaps, more convenient for values $d/l$ which fall into
the windows $0.2 < \eta < 0.3$ (for $e_3h$-a isomer) and
$1.2 < \eta < 1.5$ (for $e_3h$-b isomer): see Fig.1. Even in these windows, 
the ionization energy of $e_3h$ complexes 
amounts to only 
2-4\% of $|U_1| = e^2/(\epsilon l)$,
which means that experimental
investigations of these relatively fragile objects
would be more difficult than
investigations of the neutral $eh$ excitons. 
The efforts might be justified by a rich variety of physical phenomena in
the novel world of mobile charged electron-hole complexes.

To summarize, the analysis above demonstrates the possibility 
of the existence of a rich family of stable charged 
complexes $e_Nh$ (or $eh_N$) in the double-layer system near a metal plate. 
These mobile complexes repel each other and may exist in the ground state 
of the low-density $e$-$h$ system. 
Changing densities of the carriers gives rise to a succession of the ground 
state transformations associated with the change of relative densities of 
boson and fermion complexes.
In the case of the Bose-Einstein condensation 
of charged boson complexes,
the condensate motion would be accompanied by a {\it nonzero total} electric 
current along the QW plane.   
Experimental realization of the suggested system would be of 
considerable interest.

This work was supported in part by DFG.
The author is thankful to M. Rozman for help in computations.

\vspace{2cm}

\noindent
Figure captions 
\begin{figure}
\caption
{} 
\end{figure}

\setcounter{figure}{0}

\begin{figure}
\vspace{0.5cm}
\caption
{Energies $\{U\}$ of the classical configurations $eh$, $e_2h$, $e_3h$-a, 
and $e_3h$-b in units of $|U_1|=e^2/(\epsilon l)$ (along the vertical axis) 
as functions of $\eta = d/l$ (the horizontal axis).}
\label{fig1}
\end{figure}


\begin{references}
\bibitem{Im} ``Image charges" polarization affects only the interaction of
the particles but not their statistics. Instead of the MP one may use any
substrate with sufficiently high dielectric constant.
\end{references}
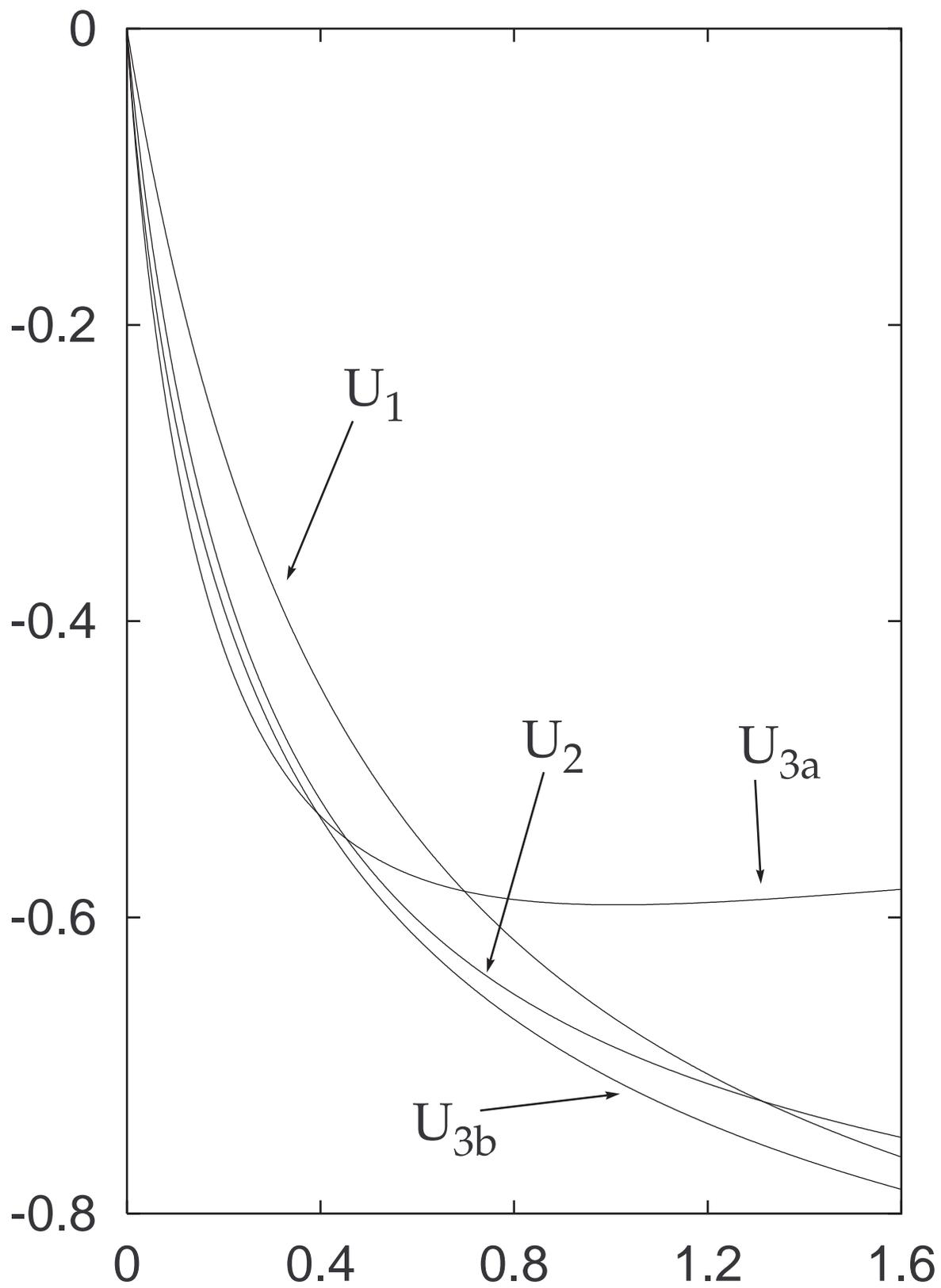
\end{document}